**Evaluating Usability in Biomedical Visualization: Rethinking Heuristic Evaluation for Spatial Omics and Multidisciplinary Research Platforms**

Yulia A. Levites Strekalova, PhD, MBA,[1*] Rachel Liu Galvin, PhD, MBChB,[1] Jessica M. Ray*,Samuel P. Border*,PhD[2], Mishal Khan, MHA[1], Samantha Hoffman, MMS, MBA[2], Christina D. Beharry, MHSc[2], Katie Kloss, BA[3], Philipp Haessner, PhD,[3] David Manthey[5], Sanjay Jain, PhD,[6] Michael T. Eadon, MD[7,8], Laura Barisoni, MD,[4] and Pinaki Sarder, PhD[2]

[1] Department of Health Services Research, Management and Policy, College of Public Health and Health Professions, Gainesville, FL, USA
[2] Department of Medicine (Quantitative Health), College of Medicine, University of Florida, Gainesville, FL, USA
[3] Department of Health Outcomes and Biomedical Informatics, College of Medicine, University of Florida, Gainesville, FL, USA
[4]Department of Pathology, Department of Medicine, Division of Nephrology, Duke University, Durham, North Carolina, USA
[5]Kitware, Inc, Clifton Park, NY, USA
[6]Department of Medicine, Division of Nephrology, Washington University School of Medicine, St. Louis, MO, USA
[7]Department of Medicine, Indiana University School of Medicine, Indianapolis, IN, USA
[8]Indianapolis VA Medical Center, Indianapolis, IN, USA

*Corresponding Author
Yulia A. Levites Strekalova, PhD, MBA
Associate Professor, Health Services Research
UF College of Public Health and Health Professions &
Director of Evaluation
UF Clinical and Translational Science Institute _
(352) 273-7934 | HPNP, 3120 | yulias@ufl.edu yulias@ufl.edu
ORCID: 0000-0002-6060-1233

**Introduction:** Clinical research informatics platforms drive biomedical discovery by integrating advanced computational tools into the research lifecycle. While emerging technologies, such as high-dimensional spatial omics and AI-powered digital imaging, expand research capabilities, their technical sophistication introduces unfamiliar interfaces that change established analytical workflows. Standard usability frameworks successfully flag surface errors but fail to capture the specific cognitive burdens imposed by dense biomedical data states.

**Methods:** To evaluate conventional usability heuristics and develop domain-specific criteria, we conducted two complementary studies with 39 participants. Study 1 engaged 19 undergraduates in interactive tasks, while Study 2 required 20 clinical professionals to complete a hierarchical task framework asynchronously. Observational and interview data were analyzed using deductive coding against standard heuristic categories and emerging CRI-specific criteria.

**Results:** Data analysis demonstrated that rendering complex overlays and analytical tools simultaneously overwhelmed users, particularly those unfamiliar with spatial-omics methods. Rather than navigating intuitively, users exhibited trial-and-error behaviors and struggled with unlabeled analytical tools in a data-rich interface. Qualitative feedback confirmed that researchers navigating higher-complexity tasks require phased onboarding and sequential tool reveals rather than immediate, default-on access to the entire feature set.

**Discussion:** High-dimensional visualization tools require tailored usability criteria. We propose three specialized heuristics: Active Parameter Transparency, Point-of-Use Guidance, and Phased Feature Disclosure. Applying these targeted heuristics ensures developers systematically introduce complexity, embed contextual help, and build accessible platforms that genuinely support multidisciplinary research.

## Introduction

Clinical research informatics (CRI) is a rapidly evolving subdiscipline of biomedical informatics that applies information science and technology across the clinical investigation lifecycle.[1,2] Recent advances in CRI have expanded the technical capabilities of clinical investigation, particularly through the emergence of AI-powered digital imaging and spatial omics platforms.[3,4] The integration of digital pathology and spatial omics is transforming biomedical image interpretation and diagnostic workflows across pathology and molecular medicine. Spatial omics technologies, in particular, are advancing the understanding of disease mechanisms and therapeutic development by preserving the microenvironmental context that traditional bulk analyses discard. Yet the technical sophistication of these platforms introduces unfamiliar interfaces and interaction demands that challenge users' existing mental models and disrupt established workflows. Systematic usability evaluation is therefore required to ensure that these tools are not only technically robust but demonstrably aligned with the needs and capabilities of their end users.

The evaluation of new CRI technologies has historically prioritized functionality and product development over usability, best practices, and human factors. As a result, many CRI tools risk misalignment with the cognitive and workflow demands of biomedical research practice.[5,6] Improving usability may reduce cognitive workload, support broader adoption across disciplines, and lower entry barriers for clinician-scientists. To realize their full potential, these tools ultimately depend on being accessible, effective, and user-friendly for a broad and diverse user base, including pathologists, oncologists, nephrologists, molecular biologists, computer scientists, and trainees across all of these disciplines.

Usability, defined as the extent to which a system enables specified users to achieve specified goals with effectiveness, efficiency, and satisfaction in a given context of use, is not merely a technical preference but a prerequisite for sustainable adoption.[7] In spatial omics tools, this requirement is especially acute. Users navigate complex visual information, interpret spatial relationships, identify relevant structures, and maintain orientation during exploratory analysis, all within interfaces that offer little margin for ambiguity. Interfaces that impose excessive cognitive burden, provide insufficient feedback, or require users to rely heavily on memory impede both accessibility and utility. High-dimensional spatial omics datasets intensify these demands. Because human working memory processes a strictly limited number of interacting elements simultaneously[8], interfaces that expose their entire feature set immediately overwhelm cognitive capacity. To maintain situation awareness, the ability to perceive and project data states[9], users require continuous visual feedback during complex data manipulation[10]. Interfaces that fail to provide this visibility impose excessive cognitive burden and impede utility. Developers of spatial omics platforms are thus tasked with addressing two simultaneous design obligations: building systems of sufficient technical power and ensuring that those systems remain interpretable and navigable for a multidisciplinary user base. Usability evaluation is the mechanism through which that balance is tested, and its role in the iterative refinement of emerging CRI tools is therefore not incidental but essential.

To systematically identify and address usability demands, heuristic evaluation serves as a well-established, expert-based method for inspecting interfaces against recognized design principles before tools reach end users.[11–13] First introduced by Nielsen and Molich in 1990[11] and subsequently refined,[12,14] The method features a set of 10 heuristics addressing system visibility, real-world alignment, user control, consistency, error prevention, recognition over recall, flexibility, aesthetic minimalism, error recovery, and help and documentation.[15,16] A key practical advantage of heuristic evaluation is its efficiency: it identifies major usability problems without the resource demands of large-scale user testing, making it well suited to iterative development cycles in research and clinical informatics contexts.[17] Evaluators typically work independently before consolidating findings, ranking identified problems by severity, frequency, and criticality, and producing a structured, actionable roadmap for interface improvement.[15]

Despite its wide adoption, heuristic evaluation carries recognized limitations. It does not involve real users, does not follow task-based workflows, and its outcomes are subject to the expertise and judgment of the evaluators selected.[16] These constraints are particularly salient in specialized or rapidly evolving technological domains, where standard heuristics developed for general-purpose interfaces may not fully capture the usability considerations most relevant to the user population.[13] Prior research confirms that Nielsen's original heuristics may require domain-specific adaptation when applied to complex health information technologies,[16] and that the adequacy of standard heuristics cannot be assumed for emerging platforms in clinical and translational research settings.[3] Specifically, multidisciplinary biomedical interfaces require interpretive scaffolding, system-embedded support that bridges knowledge gaps during active problem-solving[18,19]. They also demand progressive interface exposure, which phases in advanced functions to accelerate learning and reduce errors[20]. Grounded in cognitive theory, these specific design interventions suggest that the adequacy of standard heuristics cannot be assumed for emerging clinical research platforms. To address these limitations, combining heuristic evaluation with think-aloud interviews allows researchers to capture direct user input, reducing reliance solely on an evaluator's domain knowledge, which may be limited in novel research domains. The application of heuristic evaluation to advanced spatial omics visualization platforms, where users must simultaneously interpret high-dimensional molecular data and complex histological imagery, represents precisely this kind of specialized context and raises the open question of whether existing heuristics are sufficient or whether domain-specific criteria are warranted.

### *Objective and research question*

This study aims to demonstrate a mixed-methods framework for evaluating the usability of emerging spatial omics visualization platforms in collaborative research environments. By integrating think-aloud protocols with structured heuristic evaluation, this approach establishes a reproducible standard applicable to the broader class of high-dimensional biomedical visualization tools while examining whether conventional heuristics adequately capture the complexity of such interfaces. Two complementary studies addressed this objective: Study 1 employed think-aloud protocol interviews with students enrolled in a research experience to capture real-time interaction patterns and cognitive responses, while Study 2 applied structured heuristic evaluation via a multidisciplinary expert panel, including pathologists, molecular biologists, and trainees, to identify usability strengths and deficiencies relative to Nielsen's original heuristics. By surfacing interface deficiencies early in the software development lifecycle, this framework advances a broader principle: emerging biomedical visualization tools must be not only technically functional but demonstrably usable for their intended multidisciplinary user communities. To guide this evaluation, the study addresses two primary research questions:

> **RQ1:** To what extent do standard usability heuristics (such as Nielsen’s principles) adequately capture the unique interaction challenges, high-dimensional visualization requirements, and data-state complexities of clinical research informatics (CRI) software?
>
> **RQ2:** How can operational definitions and domain-specific criteria for usability heuristics be formulated and systematically applied to evaluate complex CRI systems within multidisciplinary research environments?

### *Technology*

Within the CRI context of spatial omics visualization, this study uses FUSION[21] as its representative case. FUSION is a cloud-based visualization and analysis platform for the integrated, interactive exploration of high-resolution whole slide images (WSIs) and spatial omics data. FUSION enables users to overlay segmentation outputs, pathomic features, and spatial omics data onto WSIs at the level of individual functional tissue units (FTUs), supporting analysis at the cell, tissue structure, and biopsy levels across a multidisciplinary user base. A key design feature that distinguishes FUSION from existing visualization tools is its "human-in-the-loop" model of interactive analysis, enabling users to iteratively refine segmentation, annotation, and feature selection in response to data observations, rather than passively viewing pre-processed outputs.[21]

## Study 1: Undergraduate Student Research Experience

## Methods

### *Setting and Participant Recruitment*

The Computational Image Analysis Platform (CIMAP) project, part of the NIH-funded Human BioMolecular Atlas Program (HuBMAP) consortium,[22] developed a five-day virtual research internship at the University of Florida in March 2024.[23] The program brought together undergraduate engineering and public health students for collaborative, interdisciplinary work in artificial intelligence and machine learning (AI/ML) applied to biomedical research. Central to the internship curriculum was evaluating the FUSION platform,[21] where participants engaged in hands-on data exploration and completed structured usability tasks to provide multi-perspective user feedback.

### *Data Collection*

All participants in the spring break research internship were enrolled in the FUSION usability study. After receiving written instructions on account setup, study materials, screen recording, and task procedures, students worked in remote pairs via Zoom to complete a guided FUSION tutorial. During these sessions, participants recorded their screens and applied a think-aloud protocol to verbalize their actions and reasoning. Following the tutorial, student pairs conducted peer interviews evaluating platform usability and potential educational applications. We supplied a standardized elicitation interview guide to ensure consistent prompting across pairs, targeting system strengths and weaknesses, essential and desired features, discipline-specific use cases, and participants' ability to articulate FUSION's core functions. The complete interview guide appears in Appendix A.

### *Data Analysis*

The rapid qualitative analysis (RQA) approach was used to analyze data. RQA is a team-based method that relies on templated summaries and matrix displays to accelerate analysis while preserving transparency.[24,25] After de-identifying and transcribing the Zoom and screen-recorded sessions, the team created a deductive template anchored to Nielsen's heuristics domains (Figure 1): (1) *Visibility of System Status* (providing timely status feedback); (2) *Match Between System and Real World* (aligning terminology and logic with user expectations); (3) *User Control and Freedom* (offering clear exit routes and error recovery); (4) *Consistency and Standards* (following established interface conventions); (5) *Error Prevention* (proactively preventing slip and mistake conditions); (6) *Recognition Rather than Recall* (minimizing memory load); (7) *Flexibility and Efficiency of Use* (supporting accelerators and custom workflows); (8) *Aesthetic and Minimalist Design* (eliminating redundant visual elements); (9) *Help Users Recognize, Diagnose, and Recover from Errors* (offering plain-language error messaging and solutions); and (10) *Help and Documentation* (providing concise, task-oriented guidance).

Figure 1. Nielsen's Ten Usability Heuristics for User Interface Design*

**1 VISIBILITY OF SYSTEM STATUS**
Keeping users **informed** of platform state through **appropriate, timely feedback.**
- Progress indicators, loading bars, system state messages

**2 MATCH BETWEEN SYSTEM AND THE REAL WORLD**
Aligning interface **concepts, language, and workflows with domain-specific conventions** and **user mental models.**
- Jargon-free language, common symbols, workflow match

**3 USER CONTROL AND FREEDOM**
Providing **clear exits and undo/redo capabilities** when actions are taken mistakenly.
- Back buttons, undo actions, safe exit routes

**4 CONSISTENCY AND STANDARDS**
Adhering to platform and industry interface **conventions** to ensure **predictable system** behavior.
- Uniform design across system, industry-standard UI elements

**5 ERROR PREVENTION**
Designing interfaces to **eliminate error-prone conditions** or present **confirmation prompts prior** to critical actions.
- Confirmation dialogs for delete, constraint verification

**6 RECOGNITION RATHER THAN RECALL**
Minimizing cognitive load by making objects, options, and actions visible rather than requiring memory retrieval.
- Examples: Visual menu lists, visible action buttons

**7 FLEXIBILITY AND EFFICIENCY OF USE**
Incorporating shortcuts, accelerators, and customizable interface elements for expert users without compromising novice usability.
- Examples: Keyboard shortcuts, custom settings, expert modes

**8 AESTHETIC AND MINIMALIST DESIGN**
Eliminating irrelevant or rarely needed visual information to preserve focus on primary task workflows.
- Examples: Focused tasks, limited elements, clear visual hierarchy

**9 HELP USERS RECOGNIZE, DIAGNOSE, AND RECOVER FROM ERRORS**
Presenting clear, jargon-free error messages that precisely indicate problems and suggest constructive solutions.
- Examples: Specific problem description, prescriptive solution guidance

**10 HELP AND DOCUMENTATION**
Supplying accessible, task-focused documentation to assist users when explicit guidance is required.
- Examples: Searchable help, context-sensitive documentation, user manuals

**Figure 1 was generated by Gemini Nano Banana 2 with the first author's prompting and review of the content*

Inductive codes were incorporated to capture emerging usability issues beyond the heuristic framework. Two analysts independently produced a templated summary for each participant pair, compared interpretations, and resolved discrepancies by consensus. The team then populated a case-by-domain matrix to support cross-case comparison, pattern detection, and prioritization of high-salience needs. Brief analytic debriefs after every three cases refined code definitions, collapsed redundancy, and tracked emergent themes in an audit trail. Analysts captured illustrative quotations and quick counts of theme occurrences to convey relative salience without over-quantification. This rapid process generated timely, credible findings suited to early-stage usability evaluation.[24,26]

### *Ethical Review*

This study was reviewed and determined to be exempt by the University of Florida Institutional Review Board (ET00054347).

## Results

### *Participants*

Twenty-one students participated in the CIMAP spring break experience, with 19 completing the think-aloud usability evaluation. Eighteen participants completed the post-program survey assessing sociodemographic and academic background characteristics (Table 1).

### *FUSION system usability*

A rapid qualitative analysis of the 19 participant interviews reveals that all 10 of Nielsen's Usability Heuristics are represented in the students' feedback, though the feedback is heavily skewed toward systemic and instructional issues. The most frequently violated heuristic is *Visibility of System Status*, widely cited due to severe server lag, lack of loading indicators, freezing screens, and a prevalent bug in which other users' inputs overwrote participants' text boxes. *Help and Documentation* was the second most discussed category; while some students found the initial PowerPoint slides helpful, many requested more integrated, contextual onboarding, such as interactive tooltips, video crash courses, or pop-up guides directly within the FUSION interface to explain complex biological features and user interface navigation. Less frequently cited, but still notable, were issues related to *User Control and Freedom* (e.g., getting trapped in zoom states or menus without clear exits), *Consistency and Standards* (e.g., dynamic slider scales and shifting color codes), and a desire for better *Error Prevention* (e.g., auto-saving responses). Conversely, the platform succeeded in *Recognition Rather Than Recall* and *Aesthetic and Minimalist Design* by automatically segmenting/identifying cells and offering visually distinct color-coding, which users found highly beneficial for rapid analysis compared to legacy software.

***RQ1: Adequacy of usability heuristics for CRI software***

Standard usability heuristics provide a useful but incomplete framework for evaluating the interaction challenges distinctive to CRI software, such as FUSION. The Study 1 think-aloud data reveal friction that standard heuristics can name but cannot fully diagnose. Several participants independently struggled to locate and activate cell-level and spatial-omics visualization features, pointing to failures that map nominally onto *Recognition Rather Than Recall* and *Visibility of System Status*. Yet, the underlying cause is more specific: the platform presents high-dimensional, multi-layered data states, including overlaid cell-type annotations, heatmaps, and composition pie charts rendered simultaneously within a single viewport, which impose a form of cognitive load for which general-purpose heuristics lack an operational vocabulary.

Participant UG-09 noted that they "couldn't figure out the heat map," and that for cell compositions and subtypes, "it took a while for it to load," underscoring that the problem is not merely a labeling deficit but an interpretive one tied to the visualization architecture itself. Participant UG-13 articulated the compounded challenge directly: "the slide itself is already lit up with so many colors when you first log on to FUSION, so that's confusion already," identifying a data-state complexity, namely default-on multi-layer overlays, that no single Nielsen heuristic addresses discretely. Similarly, participant UG-08 reported that the cell-subtype viewing feature "did not work at all in my case and I don't know if that's because I am using it wrong," a remark that conflates system transparency failure with domain knowledge deficits in ways that standard heuristics treat as separate categories but that CRI platforms collapse into a single moment of interaction.

These observations converge on a structural limitation: Nielsen's principles were designed for general-purpose software interactions and assume a relatively stable, task-bounded interface state, whereas CRI platforms like FUSION continuously alter their data-state as users navigate across tissue structures, cell types, and omics layers. Standard heuristics can flag the surface symptoms but lack the specificity to capture the domain-intrinsic complexity that generates them.

***RQ2: Formulating domain-specific heuristic criteria for CRI tools***

The Study 1 data suggest that operational criteria adequate for CRI usability evaluation must extend Nielsen's principles along at least three domain-specific axes: visualization interpretability, data-state transparency, and onboarding scaffolding for multi-disciplinary novice users. The think-aloud interviews consistently show that standard criteria for *Match Between System and the Real World* and *Help and Documentation* are necessary but insufficient. Participants needed not only conventional navigational affordances but also conceptual scaffolding for the spatial-omics data structures that the interface represents.

Participant UG-11, a biomedical engineering student, described the FUSION visualization presentation as providing "basically a whole digitized manual of the entire program," which he valued, but immediately qualified that he would need "a search bar for all the different options" and a dedicated help feature to make these materials actionable mid-task. Participant UG-06 similarly identified the absence of step-by-step procedural guidance as the primary barrier: "I kind of had to play around until I was able to find where I could actually find it. So, just more clearer structures, like, honestly going step by step almost over-explaining it would be helpful". These user-generated criteria, expressed during elicitation interviews, map onto emerging domain-specific frameworks that can reconceptualize heuristics for high-dimensional visualization environments. Participant UG-14, unable to complete the core tasks, articulated a criterion close to *progressive disclosure*: "it's a lot at once and like all on one page, there's no home page to direct you," proposing that introductory click-through walkthroughs, as used in non-scientific visualization tools like Adobe Illustrator or similar, could serve as contextual onboarding for new users. Participant UG-07 reinforced this operationally, recommending "for each feature you would have like a little pop-up screen or question mark thing on the side that can explain [to] you what it does," a criterion that functions as a domain-specific extension of *Recognition Rather Than Recall*, particularized here to the problem of unlabeled analytical tools in a data-rich interface.

Taken together, these convergent observations suggest that CRI-specific heuristics should include explicit criteria for: (1) *data-state visibility*, whether users can determine at a glance which analytical layers are active and what data are available in the current view; (2) *interpretive scaffolding*, whether the interface provides sufficient domain-contextualized guidance at the point of visualization rather than solely in external documentation; and (3) *progressive interface exposure*, whether novice users across disciplines can be systematically introduced to the platform's complexity rather than confronted with its full feature set on first access.

## Study 2: Pathologists, Biologists, and Students

## Methods

### *Setting and Participant Recruitment*

Participants were recruited from July 2023 to July 2024 via an online survey distributed across several renal pathology working groups and professional networks, including the Renal Pathology Society, the Nephrotic Syndrome Study Network, the Cure Glomerulonephropathy consortium, the International Summer School of Renal Pathology and Precision Medicine, and GlomCon, as well as during institutional grand rounds. Outreach methods included virtual and in-person presentations paired with interactive software demonstrations. Approximately 300 professional end users (pathologists, biologists, and molecular nephropathologists) were invited, of whom 20 participated.

### *Data collection*

Participants first completed an interest and demographic survey capturing professional role (for user group classification), years of experience, highest degree attained, and baseline familiarity with digital pathology, renal pathology, image analysis, and omics data interpretation. Following survey completion, participants received individualized login credentials to access the platform. To support task execution, participants were provided with dedicated tutorial slides tailored to each task level. User accounts were configured to deliver role-specific task sequences based on participant background (e.g., pathologist, biologist). All usability sessions were conducted online asynchronously. Participants were instructed to capture silent screen recordings of their system interactions during both training and task completion, subsequently uploading these recordings to a secure server. Upon video submission, participants were emailed a link to complete the System Usability Scale (SUS) usability survey.

### *Usability tasks*

Tasks were structured using a hierarchical, three-level framework codeveloped by experts in pathology, engineering, and human factors to account for varying domain expertise among pathologists, biologists, and trainees. As summarized in Table 2, Level 1 tasks assessed foundational data interactions aligned with routine discipline-specific workflows. Level 2 tasks evaluated FUSION’s utility in assisting users with complex analytical activities, such as ROI analysis, biomarker association, and sub-structure phenotyping. Finally, Level 3 tasks were identical across all cohorts, prompting participants to synthesize platform findings to generate novel, testable research hypotheses.

### *Data Analysis*

Quantitative data were summarized using descriptive statistics. Participant demographics and SUS responses were summarized using counts, percentages, means, and standard deviations. Overall usability was benchmarked using the System Usability Scale (SUS), a validated 10-item instrument scored on a 0–100 scale, where scores below 68 indicate below-average usability, 68–84 reflect above-average usability, and 85–100 signify exceptional usability.[27]

To complement the quantitative SUS benchmark and identify specific areas for system improvement, we conducted behavioral video analysis of participant sessions. Human factors researchers (YLS, JMR) and trained research assistants (KK, PH) evaluated the video recordings using continuous timestamp coding to capture observable user behaviors, exploratory actions, and reviewer observations. The analysis team regularly convened to synthesize common findings across sessions. Qualitative findings were deductively categorized using Nielsen's 10 usability heuristics framework as described in Study 1 (Figure 1).[12]

### *Ethical Review*

This study was reviewed and determined to be exempt by the University of Florida Institutional Review Board (ET00023060).

## Results

### *Participants*

Participants representing potential user groups of pathologists, biologists, and students were reached. Of approximately 300 individuals invited, 20 agreed to participate, completed a usability session, and submitted screen recordings. Demographics were collected for all 20 participants, and 14 completed the post-study usability survey (70% response rate). Table 3 displays the professional training and background experience of the study cohort. Notably, 95% of respondents reported prior experience with renal pathology, 90% with digital pathology, and 65% with image analysis. However, only 40% reported being somewhat or very familiar with interpreting omics data.

### *FUSION system usability*

**System Usability Scale.** Quantitative data from the 14 surveys were analyzed for system usability following completion of the usability tasks. Descriptive analysis indicated a mean System Usability Scale score of 54 (with individual responses ranging from 30 to 80; overall SD = 1.06). SUS results should not be interpreted as percentages but rather used as benchmarks, and the FUSION score indicated marginal acceptability.[28]

**FUSION usability heuristics.** Video analysis of the usability sessions revealed friction across six of Nielsen's ten heuristics. *Visibility of System Status* posed the most pervasive challenge. Users struggled to clear manual annotation selections, and attempts to access cell state and cell type data via the Compositions tab prompted extensive exploration. Lacking clear system feedback, participants clicked blindly to determine if they could add a cell subtype. *Recognition Rather Than Recall* also strained user interactions. Instead of relying on intuitive interface cues, participants clicked rapidly across options or returned repeatedly to task-specific tutorials to learn how to select main cell types for heatmaps, draw shapes for regions of interest, and analyze cell distributions via pie charts. *Consistency and Standards* issues clustered around the nephron diagram. Users frequently zoomed in to read small labels, and they suggested inverting text or adding global labels to prevent hover text from cutting off at the screen's right edge. Regarding the *Match between the System and the Real World*, user feedback raised concerns about pathological segmentation accuracy. One reviewer noted that the software missed some non-sclerotic and sclerotic glomeruli during visualization. *Error Prevention and Recovery* remained uneven. When opening the annotation layer menu, users accidentally toggled selections due to overlapping screen elements, though they quickly reversed these minor errors. Manual annotation errors, by contrast, triggered complex recovery attempts as users struggled to clear erroneous areas. Finally, *Help and Documentation* served as a necessary crutch. While some participants scanned the training slides quickly, others relied heavily on the tutorials, returning to them repeatedly because the interface lacked intuitive guidance. To bridge this gap, users requested embedded video tutorials and additional training for interpreting complex data visualizations.

### *RQ1: Adequacy of usability heuristics for CRI software*

Standard usability heuristics provide a useful but incomplete framework for evaluating the interaction challenges distinctive to CRI software, such as FUSION. The Study 1 think-aloud data reveal friction that standard heuristics can name but cannot fully diagnose. Likewise, professionals who participated in Study 2 struggled with the same visualization features that were challenging for the undergraduate cohort. Video data captured Participant 19 repeatedly leaving the main interface to review task-specific tutorial slides, explicitly seeking guidance on configuring spatial omics and heatmap overlays. Another expert user, P-12, spent several minutes rapidly toggling cell types while viewing the heatmap in a blind trial-and-error search. While traditional heuristics might categorize these behaviors as failures in *Help and Documentation* or *Error Recovery*, the root issue remains systemic. The interface fails to clarify how complex, layered inputs translate into visual outputs.

These observations converge on a structural limitation: Nielsen's principles were designed for general-purpose software interactions and assume a relatively stable, task-bounded interface state. By contrast, CRI platforms like FUSION continuously alter their data-state parameters as users navigate across tissue structures, cell types, and omics layers. Standard heuristics successfully flag the general navigation issues, but they lack the specificity to capture the domain-intrinsic complexity that generates them.

### *RQ2: Formulating domain-specific heuristic criteria for CRI tools*

To address how operational definitions for CRI heuristics can be formulated, video data from Study 2 were analyzed to determine if the usability friction observed in Study 1 persisted among experienced professionals. The findings indicate that domain expertise did not resolve the structural interface challenges, supporting the formulation of three suggested CRI-specific heuristics.

First, *data-state visibility* requires interfaces to show active analytical layers at a glance. Because the system lacked clear indicators for activated cell subtypes, Participant 12 spent several minutes toggling heatmap inputs in a trial-and-error search, and multiple users struggled to clear manual annotations without visual confirmation of their tool state. Second, *interpretive scaffolding* demands contextual guidance directly within the visualization rather than in siloed documentation. Separated training materials disrupted workflows; Participant 11 paused analysis twice to read every tutorial slide. Users explicitly requested localized help, noting "the tools needed to have more description on what it does," and some resorted to external web searches to interpret heatmaps. Finally, *progressive interface exposure* prevents cognitive overload by introducing complexity systematically. Rendering complex overlays and analytical tools simultaneously overwhelmed users, especially those unfamiliar with spatial-omics methods. Qualitative feedback confirmed that researchers navigating higher-complexity tasks require phased onboarding and sequential tool reveals rather than immediate, default-on access to the entire feature set.

## Discussion

### *Summary of the present studies*

This study presents the initial usability analysis and benchmarking of FUSION, a spatial omics visualization platform developed for multidisciplinary clinical and research use. Results indicated marginally acceptable usability based on the System Usability Scale (SUS) benchmarks, alongside behavioral observations that revealed specific opportunities for interface improvement. Observed confusion with task completion may reflect both the novelty of spatial omics interfaces and ambiguities in task wording, underscoring the well-established importance of involving end users across expertise levels in task development for usability studies. The marginal SUS score of 54 falls below the commonly cited threshold of 68 for average usability; future iterations will target a score of 70 or above, consistent with the usability benchmarks of widely adopted productivity tools.

Our findings suggest that some users were confused by the tasks themselves. This could be an artifact of users' experience with the system's spatial omics components or of the wording of the tasks. While our tasks were developed by a team that included content experts, future work should engage multiple users with varying levels of expertise in task development to ensure the tasks themselves do not become part of the system usability assessment. Similarly, additional work is needed as system improvements are implemented to ensure that the tutorials and system documentation remain relevant and usable by the end user. While the marginally acceptable SUS rating for FUSION is not what we had hoped, it does provide a baseline. To reach an acceptable level, our future iterations will look to achieve a score of 70 or better, which would align the tool with non-scientific applications such as Microsoft Word or Adobe Illustrator.[29]

Despite being a critical step in the development of any new system, formal usability evaluation remains underutilized in academic and research informatics contexts relative to commercial software development. Barriers to recruiting appropriate and diverse user groups present a persistent challenge in this work. To address this, the present study employed a remote, asynchronous testing approach, demonstrating one scalable method for engaging large, geographically distributed user communities across healthcare and academic settings.

***Limitations and strengths***

The remote, asynchronous design introduced several methodological limitations. Data captured were predominantly behavioral; think-aloud protocols or follow-up interviews would have provided richer insight into users' cognitive processes and affective responses during task completion, a recognized limitation of remote testing formats. Additionally, technical issues with the task presentation and answer-entry system, unrelated to FUSION itself, could not be corrected in real time due to the asynchronous format, potentially introducing noise into the usability data. Future studies should consider hybrid designs that combine asynchronous behavioral data collection with synchronous debriefing sessions to mitigate these constraints.

The geographic and disciplinary diversity of participants, while contributing to variability in usability scores, represents a key methodological strength. Remote, asynchronous testing enabled engagement of participants across multiple countries, a recruitment scope rarely achievable in traditional synchronous sessions, given international scheduling constraints. This breadth enhances the ecological validity of the findings and supports the generalizability of the usability assessment to FUSION's intended multidisciplinary, global user base.

***Alignment with existing literature***

The proposed heuristics translate foundational cognitive theories into actionable evaluation criteria for contemporary clinical research informatics. Data-state visibility adapts Endsley's theory of situation awareness[9] and visual feedback principles[10] for multi-layered omics viewports. By demanding explicit visual confirmation of active analytical layers, this metric ensures users comprehend their data environment before executing complex tasks. Recent usability evaluations of AI-integrated medical imaging reinforce this requirement, demonstrating that dynamic state-tracking prevents user disorientation during complex diagnostic workflows [3]. Similarly, interpretive scaffolding shifts help features from passive documentation to active cognitive support. Drawing on educational psychology[18] and learner-centered design[19], this criterion requires systems to embed contextual guidance directly at the point of visualization. Current informatics research validates this approach; recent usability studies show that point-of-use guidance significantly lowers interpretative error rates in high-dimensional biomedical platforms[30]. Finally, progressive interface exposure mitigates working memory limits[8]. Building on the concept of progressive disclosure models[20], this heuristic compels developers to phase in advanced tools systematically. Furthermore, as contemporary literature argues, modern evaluation frameworks must address the extreme density of contemporary software[15]. Phasing complexity prevents the cognitive overload that occurs when systems confront multidisciplinary users with default-on visual overlays[21]. Together, these metrics move usability assessment beyond surface navigation. They target the structural cognitive demands of science itself.

To evaluate complex clinical research platforms effectively, developers must apply these theoretical axes as practical design constraints. First, data-state visibility adapts traditional system status principles for high-dimensional environments, compelling systems to indicate active analytical layers at a glance. This requirement prevents the disorientation that results when multi-layered viewports obscure tool states. Second, interpretive scaffolding transforms basic documentation into embedded, contextual guidance. Instead of forcing users to consult disconnected training materials, platforms must dynamically bridge knowledge gaps directly at the point of visualization. Finally, progressive interface exposure curtails the cognitive overload caused by simultaneous feature rendering. Rather than confronting multidisciplinary teams with a complete toolset upon initial access, software architecture should introduce analytical complexity systematically. By integrating these three criteria, evaluators align heuristic assessments with the rigorous, layered demands of modern biomedical research.

***Implications for future research and practice***

The three CRI-specific heuristics proposed here, data-state visibility, interpretive scaffolding, and progressive interface exposure, offer a transferable evaluation framework that future researchers could test across other high-dimensional biomedical visualization platforms. This study examined a single platform with a relatively small sample, and the variance in System Usability Scale scores signals that FUSION's usability profile shifts substantially across disciplinary backgrounds, expertise levels, and omics familiarity. Future longitudinal studies that track the same users across iterative interface releases should clarify whether targeted design interventions, such as progressive tool reveals and embedded contextual guidance, produce measurable reductions in cognitive load over time. Equally important, future heuristic evaluation work should involve domain experts as co-evaluators from the outset, rather than applying criteria developed for general-purpose interfaces retrospectively. Extending the present framework to platforms in genomics, proteomics, and AI-assisted radiology would test whether the three heuristics generalize to high-dimensional visualization environments beyond spatial omics, or whether additional domain-specific criteria are needed.

For developers and clinical research teams deploying spatial omics tools in practice, the findings carry immediate design mandates. First, default interface states should expose only foundational analytical layers, reserving advanced overlays, such as cell-subtype heatmaps and multi-channel composition plots, for users who have completed structured onboarding sequences. Confronting multidisciplinary users with the full feature set on first access reliably triggers the trial-and-error behavior documented in both studies, wasting clinical time and undermining confidence in the platform. Second, help resources should migrate from external documentation into the interface itself, appearing at the precise point where a user encounters an unfamiliar tool or data state. This shift transforms support from a passive reference into an active scaffold, as educational psychology has long demonstrated is necessary for bridging knowledge gaps during active problem-solving. Finally, as the pathology workforce faces increasing demand for computationally integrated assessments that extend well beyond traditional visual scoring, institutions that treat usability evaluation not as a downstream quality-control step but as a core component of the development lifecycle. Tools that are technically robust but demonstrably unusable will not achieve the adoption necessary to realize the translational potential of spatial omics technologies.

**Table 1. Participant Demographics, Study 1**

| Characteristics | Percent | N |
|---|---|---|
| *Sex* | | |
| Male | 35% | 6 |
| Female | 65% | 11 |
| *Race* | | |
| White | 39% | 7 |
| Black or African American | 28% | 5 |
| Asian | 28% | 5 |
| Other | 6% | 1 |
| *Ethnicity* | | |
| Hispanic or Latino | 29% | 5 |
| Not Hispanic or Latino | 71% | 12 |
| *Highest level of parents' education* | | |
| High school or less | 12% | 2 |
| Associate's degree | 18% | 3 |
| College degree | 47% | 8 |
| Advanced or professional degree (PhD, MD, etc.) | 24% | 4 |
| *Grew up in a rural area* | | |
| Yes | 18% | 3 |
| No | 82% | 14 |
| *Class standing* | | |
| Sophomore | 18% | 3 |
| Junior | 35% | 6 |
| Senior | 47% | 8 |
| *Major* | | |
| Biomedical Engineering | 47% | 8 |
| Public Health | 41% | 7 |
| Electrical Engineering | 12% | 2 |

[a]Percentages may not sum to 100% due to rounding. Sample sizes vary slightly across variables (n = 17–18) because of item-level nonresponse.

**Table 2. Usability tasks**

| Task Level | Target User | Task |
|---|---|---|
| One | Pathologist | Record basic observations of tissue morphology and structure counts |
| | Biologist | Assess cellular localization across the same sample of tissue |
| | Trainee | Identify differences between structures |
| Two | Pathologist | Determine the presence of cell types in different types of structures (including manual regions of interest [ROIs] containing lesions) |
| | Biologist | Associate cell types with structures and associate injury biomarkers with physical evidence of injury |
| | Trainee | Identify sub-categories of structures based on either cell type composition or phenotype |
| Three | Pathologist<br>Biologist<br>Trainee | Expand on the data presented directly in FUSION to formulate hypotheses that could be tested either within FUSION or through secondary analysis |

**Table 3. Participant Demographics, Study 2**

| Category | Percent | N |
|---|---|---|
| *Professional role* | | |
| Pathologist | 85% | 17 |
| Graduate Student | 5% | 1 |
| Nephrologist | 5% | 1 |
| Basic Scientist | 5% | 1 |
| *Highest degree* | | |
| MD | 65% | 13 |
| MD/PhD | 15% | 3 |
| PhD | 5% | 1 |
| Masters | 5% | 1 |
| Bachelors | 5% | 1 |
| Other* | 5% | 1 |
| *Years of experience (for non-trainee participants only)* | | |
| 5 or less | 15% | 3 |
| 6-10 | 20% | 4 |
| 11-15 | 5% | 1 |
| 16-20 | 30% | 6 |
| 21-25 | 20% | 4 |
| Over 25 | 0% | 0 |
| N/A | 12% | 2 |
| *Experience with digital pathology* | | |
| Yes | 90% | 18 |
| No | 10% | 2 |
| *Experience with renal pathology* | | |
| Yes | 95% | 19 |
| No | 5% | 1 |
| *Experience with image analysis* | | |
| Yes | 65% | 13 |
| No | 35% | 7 |
| *Familiarity with omics data* | | |
| Very unfamiliar | 30% | 6 |
| Somewhat unfamiliar | 10% | 2 |
| Neither familiar nor unfamiliar | 20% | 4 |
| Somewhat familiar | 30% | 6 |
| Very Familiar | 10% | 2 |
| *Familiarity with image analysis software (e.g. QuPath, Aperio ImageScope, etc.)* | | |
| Very unfamiliar | 5% | 1 |
| Somewhat unfamiliar | 10% | 2 |
| Neither familiar nor unfamiliar | 5% | 1 |
| Somewhat familiar | 55% | 11 |
| Very Familiar | 25% | 5 |

*One participant selected "Other" for highest degree and entered "Fellow of the Royal College of Physicians and Surgeons of Canada," which is a professional fellowship rather than an academic degree. This response was retained as reported.

**Table 4. CRI Heuristics**

| Heuristic | Definition | Example | Application to CRI |
|---|---|---|---|
| *Visibility of System Status* | The system should always keep users informed about what is going on through appropriate feedback within a reasonable time. | A progress bar while a file is uploading. | |
| *Match Between System and the Real World* | The system should speak the user's language, using words, phrases, and concepts familiar to the user, rather than system-oriented terms. | Using a trash can icon for "Delete" rather than a cryptic technical command. | |
| *User Control and Freedom* | Users often perform actions by mistake and need a clearly marked "emergency exit" to leave the unwanted state without having to go through an extended dialogue. | Providing an "Undo" button or a clear "Cancel" option. | |
| *Consistency and Standards* | Users should not have to wonder whether different words, situations, or actions mean the same thing. Follow platform conventions. | Keeping the "Settings" menu in the same location across an entire application. | |
| *Error Prevention* | Even better than good error messages is a careful design that prevents a problem from occurring in the first place. | Graying out a "Submit" button until all mandatory fields are filled out. | |
| *Recognition Rather Than Recall* | Minimize the user's memory load by making objects, actions, and options visible. The user should not have to remember information from one part of the interface to another. | Displaying recently viewed items instead of forcing the user to search for them again. | |
| *Flexibility and Efficiency of Use* | Accelerators—unseen by the novice user—may often speed up the interaction for the expert user. | Keyboard shortcuts (like Ctrl+C for copy) for power users. | |
| *Aesthetic and Minimalist Design* | Interfaces should not contain information that is irrelevant or rarely needed. Every extra unit of information in an interface competes with the relevant units of information. | Removing decorative elements that clutter the screen and distract from the primary task. | |
| *Help Users Recognize, Diagnose, and Recover from Errors* | It is best if the system can be used without documentation, but it may be necessary to provide help and documentation that is easy to search and focused on the user's task. | A contextual "Help" or "FAQ" link placed near a complex feature. | |

**References**

1. Embi PJ, Payne PRO. Clinical Research Informatics: Challenges, Opportunities and Definition for an Emerging Domain. *J Am Med Inform Assoc*. 2009;16(3):316-327. doi:10.1197/jamia.M3005
2. Kahn MG, Weng C. Clinical research informatics: a conceptual perspective. *J Am Med Inform Assoc*. 2012;19(e1):e36-e42. doi:10.1136/amiajnl-2012-000968
3. Embi PJ. Clinical Research Informatics: Survey of Recent Advances and Trends in a Maturing Field. *Yearb Med Inform*. 2013;22(1):178-184. doi:10.1055/s-0038-1638853
4. Ohno-Machado L. Clinical research informatics: a growing subspecialization of biomedical informatics. *J Am Med Inform Assoc JAMIA*. 2018;25(3):223. doi:10.1093/jamia/ocy008
5. Improving Care: Priorities to Improve Electronic Health Record Usability.
6. Taft T, Staes C, Slager S, Weir C. Adapting Nielsen's Design Heuristics to Dual Processing for Clinical Decision Support. *AMIA Annu Symp Proc*. 2017;2016:1179-1188.
7. ISO 9241-11:2018. ISO. Accessed May 11, 2026. https://www.iso.org/standard/63500.html
8. Sweller J. Cognitive load during problem solving: Effects on learning. *Cogn Sci*. 1988;12(2):257-285. doi:10.1016/0364-0213(88)90023-7
9. Endsley MR. Toward a Theory of Situation Awareness in Dynamic Systems. *Hum Factors*. 1995;37(1):32-64. doi:10.1518/001872095779049543
10. Shneiderman B. The Eyes Have It: A Task by Data Type Taxonomy for Information Visualizations. In: Bederson BB, Shneiderman B, eds. *The Craft of Information Visualization*. Interactive Technologies. Morgan Kaufmann; 2003:364-371. doi:10.1016/B978-155860915-0/50046-9
11. Nielsen J, Molich R. Heuristic evaluation of user interfaces. In: *Proceedings of the SIGCHI Conference on Human Factors in Computing Systems*. CHI '90. Association for Computing Machinery; 1990:249-256. doi:10.1145/97243.97281
12. Nielsen J. Heuristic evaluation. In: *Usability Inspection Methods*. John Wiley & Sons, Inc.; 1994:25-62.
13. Gonzalez-Holland E, Whitmer D, Moralez L, Mouloua M. Examination of the Use of Nielsen's 10 Usability Heuristics & Outlooks for the Future. *Proc Hum Factors Ergon Soc Annu Meet*. 2017;61(1):1472-1475. doi:10.1177/1541931213601853
14. Nielsen J. Finding usability problems through heuristic evaluation. In: *Proceedings of the SIGCHI Conference on Human Factors in Computing Systems*. CHI '92. Association for Computing Machinery; 1992:373-380. doi:10.1145/142750.142834
15. Gonzalez Capdevila M, Pistili Rodrigues KA, Sartório Furlan TA, Granollers T. Quantifying heuristic evaluation. *Comput Stand Interfaces*. 2025;92:103891. doi:10.1016/j.csi.2024.103891
16. Zaharias P, Koutsabasis P. Heuristic evaluation of e-learning courses: a comparative analysis of two e-learning heuristic sets. *Campus-Wide Inf Syst*. 2011;29(1):45-60. doi:10.1108/10650741211192046
17. Abulfaraj A, Steele A. Coherent Heuristic Evaluation (CoHE): Toward Increasing the Effectiveness of Heuristic Evaluation for Novice Evaluators. In: Marcus A, Rosenzweig E, eds. *Design, User Experience, and Usability. Interaction Design*. Springer International Publishing; 2020:3-20. doi:10.1007/978-3-030-49713-2_1
18. Wood D, Bruner JS, Ross G. The role of tutoring in problem solving. *J Child Psychol Psychiatry*. 1976;17(2):89-100. doi:10.1111/j.1469-7610.1976.tb00381.x
19. Quintana C, Reiser BJ, Davis EA, et al. A Scaffolding Design Framework for Software to Support Science Inquiry. *J Learn Sci*. 2004;13(3):337-386. doi:10.1207/s15327809jls1303_4
20. Carroll JM, Carrithers C. Training wheels in a user interface. *Commun ACM*. 1984;27(8):800-806. doi:10.1145/358198.358218

21. Border SP, Ferreira RM, Lucarelli N, et al. FUSION: a web-based application for in-depth exploration of multi-omics data with brightfield histology. *Nat Commun*. 2025;16(1):8388. doi:10.1038/s41467-025-63050-9
22. Jain S, Pei L, Spraggins JM, et al. Advances and Perspectives for the Human BioMolecular Atlas Program (HuBMAP). *Nat Cell Biol*. 2023;25(8):1089-1100. doi:10.1038/s41556-023-01194-w
23. Liu-Galvin R, Atchison L, Ray JM, et al. An Innovative Pilot Program Approach to Facilitating Interdisciplinary Collaboration Among STEM and Public Health Students in Biomedical AI and Clinical Translational Research. *Biochem Mol Biol Educ*. 2025;53(6):666-676. doi:10.1002/bmb.70016
24. Hamilton AB, Finley EP. Reprint of: Qualitative methods in implementation research: An introduction. *Psychiatry Res*. 2020;283:112629. doi:10.1016/j.psychres.2019.112629
25. Vindrola-Padros C. Doing rapid qualitative research. Published online 2021.
26. Nevedal AL, Reardon CM, Opra Widerquist MA, et al. Rapid versus traditional qualitative analysis using the Consolidated Framework for Implementation Research (CFIR). *Implement Sci*. 2021;16(1):67. doi:10.1186/s13012-021-01111-5
27. Brooke J. SUS - A quick and dirty usability scale.
28. Vlachogianni P, Tselios N. Perceived usability evaluation of educational technology using the System Usability Scale (SUS): A systematic review. *J Res Technol Educ*. 2022;54(3):392-409. doi:10.1080/15391523.2020.1867938
29. Kortum PT, Bangor A. Usability Ratings for Everyday Products Measured With the System Usability Scale. *Int J Hum-Comput Interact*. 2013;29(2):67-76. doi:10.1080/10447318.2012.681221
30. Dejen GM. Multimodal artificial intelligence in medical biotechnology: Integrating genomics, imaging, and clinical data for precision therapeutics. *Precis Med Sci*. n/a(n/a). doi:10.1002/prm2.70047